\documentclass[journal,twocolumn]{IEEEtran}
\usepackage{epsfig,makeidx,color}
\usepackage{amsmath,amssymb,bbm}
\usepackage{cite,graphicx,lipsum}
\usepackage{enumerate}
\usepackage[switch,pagewise]{lineno}
\usepackage{hyperref}
\hypersetup{
        colorlinks = true,
        citecolor=blue,
}
\def\cK{{\cal K}}

\def\rH{{\rm H}}
\def\rT{{\rm T}}

\def\uR{{\mathbb R}}
\def\uC{{\mathbb C}}
\def\uE{{\mathbb E}}

\DeclareMathOperator*{\argmin}{\arg\!\min}

\def\be{ \begin{equation} }
\def\ee{ \end{equation} }
\def\bea{ \begin{eqnarray} }
\def\eea{ \end{eqnarray} }

\def\bx{{\bf x}}

\def\bc{{\bf c}}

\def\bb{{\bf b}}

\def\bg{{\bf g}}

\def\ba{{\bf a}}
\def\br{{\bf r}}

\def\bm{{\bf m}}
\def\bn{{\bf n}}
\def\bh{{\bf h}}

\def\bp{{\bf p}}
\def\bz{{\bf z}}

\def\bee{{\bf e}}

\def\bv{{\bf v}}
\def\bw{{\bf w}}

\def\bI{{\bf I}}

\def\bN{{\bf N}}

\def\bR{{\bf R}}

\def\b0{{\bf 0}}

\def\bSigma{{\bf \Sigma}}

\def\cC{{\cal C}}

\def\cB{{\cal B}}
\def\cP{{\cal P}}

\def\cS{{\cal S}}

\def\cN{{\cal N}}

\def\sMSE{{\sf MSE}}

\ifCLASSOPTIONonecolumn
  \newcommand{\figwidth}{0.40\columnwidth}
  
\else
  \newcommand{\figwidth}{0.85\columnwidth}
  
\fi

\begin{document}

\title{Communication-Efficient Distributed
SGD using Preamble-based Random Access}

\author{Jinho Choi\\
\thanks{The author is with
the School of Information Technology,
Deakin University, Geelong, VIC 3220, Australia
(e-mail: jinho.choi@deakin.edu.au).
This research was supported
by the Australian Government through the Australian Research
Council's Discovery Projects funding scheme (DP200100391).}}

\maketitle
\begin{abstract}
In this paper, we study communication-efficient
distributed stochastic gradient descent (SGD) with data sets 
of users distributed over a certain area and communicating
through wireless channels.
Since the time for one iteration in the proposed approach 
is independent of the number of users,
it is well-suited to scalable distributed SGD.
Furthermore, since the proposed approach 
is based on preamble-based random access, which is widely
adopted for machine-type communication (MTC),
it can be easily employed for training models with
a large number of devices in various
Internet-of-Things (IoT) 
applications where MTC is used for their connectivity.
For fading channel, we show that noncoherent combining
can be used. As a result, no channel state information (CSI)
estimation is required.
From analysis and simulation results,
we can confirm that the proposed approach
is not only scalable, but also provides improved performance
as the number of devices increases.
\end{abstract}


\ifCLASSOPTIONonecolumn
\baselineskip 28pt
\fi

\section{Introduction}

Machine learning \cite{Bishop06} \cite{Goodfellow16} has 
advanced over the past decades and 
its application has grown significantly.
For machine learning with distributed computing power and/or data sets,
distributed machine learning \cite{Verbraeken20}
has also been extensively studied,
which would play a key role in the Internet-of-Things (IoT)
where distributed devices and sensors collect and generate 
data sets.

As a distributed machine learning
approach, federated learning
\cite{FO16} \cite{Yang19} has been extensively studied,
in which users do not need to send their data sets
to a server for data privacy.
In federated learning, distributed stochastic gradient descent
(SGD) \cite{Bottou18} is used 
to update the parameter vector of a certain objective function.
In particular,
each user uploads its local update
with own data sets to a server
and the server sends the updated parameter vector
back to users for the next iteration.
Through iterations, 
the server is able to train its model with 
data sets of users without having them.
There are a number of applications of federated learning.
For example, in \cite{Pokhrel20}, for the Internet-of-Vehicles (IoV),
a federated learning framework
was proposed with stabilized data flow dynamics.
Interestingly, it is shown in \cite{Samarakoon20} that 
federated learning can also be used for 
vehicular networks to support reliable low-latency communications.

As demonstrated in federated learning, distributed SGD
is a key tool to learn the model with distributed data sets
and can be used for a number of applications where 
users with data sets are physically distributed over a certain area
and connected through wireless channels
such as mobile phones in cellular systems
or devices in IoT networks
\cite{Savazzi20}.
While there are a number of advantages of distributed SGD,
there are also challenges.
In particular, when distributed SGD is considered
for wireless applications, the scalability
becomes a critical issue as the bandwidth of wireless channels is limited.
To mitigate this problem, 
in \cite{Amiri20} \cite{Zhu20} \cite{Yang20},
the notion of over-the-air
computation \cite{Nazer07} \cite{Goldenbaum13}
is adopted for users so that 
they upload local gradient vectors simultaneously in distributed SGD.
Then, the receiver, which is a base station (BS) or access point (AP),
receives an aggregation of users' local updates by the superposition
nature of radio communication,
which leads to communication-efficient
distributed SGD.

Since machine-type communication (MTC)
becomes popular to support massive connectivity 
for a large number of devices \cite{Bockelmann16}
\cite{Ding_20Access},
it would be desirable for distributed SGD 
to use MTC protocols 
in wireless applications
with distributed devices' data sets
(hereafter, we will use the terms users and devices interchangeably).
In \cite{Choi20WCL}, distributed SGD is studied with
multichannel ALOHA that has been used 
to design a number of MTC protocols.

In this paper, we focus on distributed SGD
with distributed devices that have local data sets for training.
In principle, the model 
in this paper is the same as that in \cite{Yang20}.
Since the approach in \cite{Yang20} requires the 
channel state information (CSI) at the receiver or BS,
devices need to send pilot signals to allow the 
BS to estimate the CSI, 
which limits the scalability. That is, with a limited bandwidth
or uploading time, the number of devices participated in updating
per iteration
becomes limited.
To avoid this problem, we propose a different approach
that is based on the notion of random access. In particular,
a typical random access approach used in MTC based on 
preamble transmissions \cite{Kim20}
\cite{Choi21} is adopted so that distributed SGD can be 
implemented with MTC protocols. 

The contributions of the paper are summarized as follows: 
\begin{enumerate}
\item A communication-efficient approach to upload local
gradient vectors is proposed as a random access scheme
where the time for one iteration is independent of the number of devices;

\item To avoid the CSI estimation at the BS for
fading channels, the proposed
approach includes noncoherent combining and the 
asymptotic performance is analyzed (and compared with
the approach in \cite{Yang20});

\item To reduce the 
mean squared error (MSE)
of the proposed approach, we also consider
a modification by dividing a gradient vector 
into multiple subvectors and encoding them independently.
\end{enumerate}

It is noteworthy that 
the receiver or BS in
the proposed approach receives a superposition of 
signals transmitted by a large number of devices using random access.
Furthermore, devices do not send their identification
sequences when transmitting their updates.
As a result, the server does not know individual update or information
transmitted by any specific device. From this,
data privacy is naturally preserved,
which is another salient feature of the proposed approach.
However,
we will not discuss privacy issues any further in this paper
as we mainly focus on the communication efficiency
of distributed SGD.

Another important difference from \cite{Amiri20} \cite{Yang20}
is that the proposed approach does not rely on analog transmission
for over-the-air computation.
As stated earlier, the proposed approach is based
on a random access scheme using a set of preambles in MTC,
which means that, in principle, it is based on digital transmission
(as a local gradient vector is to be quantized).
As a result, the proposed approach is free from any
radio frequency (RF) circuit impairment.
Furthermore, unlike the approaches in 
\cite{Amiri20} \cite{Yang20}, no CSI is used at devices.
In \cite{Amiri20}, opportunistic transmissions by devices
are considered by taking into account CSI,
while the BS of the approach in \cite{Yang20} needs
to feed the CSI back to devices.
Since the proposed approach in this paper uses noncoherent combining
over fading channels, 
it is not necessary for devices to know their CSI.

The rest of the paper is organized as follows.
In Section~\ref{S:SM}, we present the system model for distributed
SGD. The quantization approach in
\cite{Gandikota19} is explained and modified in Section~\ref{S:Quan},
which is then combined with preamble-based random access in
Section~\ref{S:AWGN} 
under the additive white Gaussian channel (AWGN) model
where no CSI estimation is required.
We extend the proposed approach over fading channels
in Section~\ref{S:Fad},
and compare it with the approach
in \cite{Yang20} in Section~\ref{S:Yang}.
We present simulation results in Section~\ref{S:Sim} and conclude the
papers with remarks in Section~\ref{S:Con}.

\subsubsection*{Notation}
Matrices and vectors are denoted by upper- and lower-case
boldface letters, respectively.
The superscript $\rT$ and $\rH$
denotes the transpose and Hermitian transpose, respectively.
Denote by $||\bx||_p$ the $p$-norm of $\bx$.
For convenience, let
$||\bx|| = ||\bx||_2$, i.e., if there is no subscript, it is the 2-norm.
$\uE[\cdot]$
and ${\rm Var}(\cdot)$
denote the statistical expectation and variance, respectively.
$\cN(\ba, \bR)$
and $\cC \cN(\ba, \bR)$
represent the distributions of
Gaussian and
circularly symmetric complex Gaussian (CSCG)
random vectors with mean vector $\ba$ and
covariance matrix $\bR$, respectively.

\section{System Model}	\label{S:SM}

In this section, we present the system model 
consisting $K$ devices
with local data sets and one BS or AP. 
Here, devices and a BS 
can be regarded as compute nodes 
(or workers) and a parameter server,
respectively,
in the context of distributed machine learning.


Suppose that the BS 
wants to find the parameter vector, denoted by $\bw \in \uR^L$, 
where $L$ is the length of $\bw$,
that minimizes
a cost function, e.g.,
\be
\bw^* 
= \argmin_\bw f(\bw) 
= \argmin_\bw \frac{1}{K} \sum_{k=1}^K C( \bw; \bx_k),
\ee
where $C(\cdot)$ represents a cost function,
$\bx_k$ denotes the $k$th data set,
and $K$ is the number of data sets.
Throughout the paper, we assume
data sets are distributed over $K$ devices.
In particular, device $k$ has the $k$th data set, 
$\bx_k$.

Devices may not want to send their data sets due to data privacy
issues.
Thus, as in federated learning
\cite{FO16} \cite{Yang19},
the BS sends 
the parameter vector to the devices, and
the devices 
update the parameter vector with their data sets
using distributed SGD and send back to the BS.
Then, without sending data sets by devices,
the BS 
is able to obtain the optimized parameter vector
through iterations. 

In distributed SGD, the  
BS is  to choose one or multiple
devices uniformly at random at each iteration/round. 
In this paper, we assume that one slot is used
for one iteration.
That is, at the beginning of a slot,
the BS broadcasts the previous parameter vector.
Then, within the slot,
the selected devices compute their local gradient vectors
and upload them to the BS.
For convenience, 
we ignore the time for computing local gradient vector
and assume that the duration of one slot is mainly used to
upload local gradient vectors by devices.
Denote by $\bw_{t}$ 
the parameter vector updated at the end of slot $t$. 
Let $\cK (t)$ denote the index of the selected devices
in slot $t$,
which are chosen from $\{1, \ldots, K\}$, uniformly at random.
Then, for (minibatch) SGD \cite{Bottou18},
the updating rule at the BS is as follows:
\be
\bw_t = \bw_{t-1} - \mu \hat \bg_t,
	\label{EQ:sgd}
\ee
where the estimated aggregation of the gradient
vectors, $\hat \bg_t$, is given by
\be
\hat \bg_t = 
\frac{1}{|\cK(t)|} \sum_{k \in \cK(t)} \nabla C (\bw_{t-1}, \bx_{k}).
	\label{EQ:SGD2}
\ee
Here, $\nabla C (\bw_{t-1}, \bx_{k})$ denotes
the local gradient vector at device $k$, $\mu$ is the step size,
and $\bar K = |\cK(t)|$ is the size of minibatch.
In this paper, for convenience, we assume that 
the size of minibatch is fixed for any iteration.

Since $\cK(t)$ is a subset of $\cK = \{1, \ldots, K\}$
that are chosen uniformly at random,
it can be shown that
\be
\bg_t = \uE[\hat \bg_t] = \frac{1}{K} 
\sum_{k=1}^K \nabla C  (\bw_{t-1}, \bx_{k}),
\ee
which shows that $\uE[\bw_t]$ 
($= \uE[\bw_{t-1}] - \mu \bg_t$) follows the 
updating rule of the conventional gradient descent algorithm.

There are two key performance metrics:
\emph{i)} the time for one iteration or round (or the length
of slot);
\emph{ii)} the MSE 
of the estimated aggregation, $\uE[||\hat \bg_t - \bg_t||^2]$,
that decides 
the size of the noise ball in steady-state
(or the steady-state MSE of the parameter vector, 
$\uE[||\bw^* - \bw_t||^2]$ as $t \to \infty$).
The former metric is related to the scalability
of distributed SGD. 
Without having any parallel channels, we expect
that the time for one iteration is proportional 
to the size of minibatch, $\bar K$.
The latter metric is usually inversely proportional to
$\bar K$ \cite{Bottou18} \cite{Bernstein18}.
As a result, we face a dilemma in which 
$\bar K$ cannot be increased or decreased.

Fortunately, in this paper, we will show that the proposed approach
can avoid this dilemma such that the size of minibatch
can be the maximum (i.e., $K$), without increasing
the communication cost (i.e., the time for one iteration
is fixed regardless of $K$).


\section{Quantization for Parameter Updating}	\label{S:Quan}

In \cite{Alistarh17} \cite{Bernstein18} \cite{Gandikota19},
each device computes its local gradient vector and
quantizes it for encoding, and sends the encoded 
one to the BS. In this section, we briefly discuss the approach in 
\cite{Gandikota19},
because it is well-suited to the proposed
approach using
preamble-based random access in this paper
(we will explain this later).

\subsection{Vector Quantization using Convex Combination}

For convenience, we omit the device index $k$.
Let $\bv \in \uR^L$ be a vector,
which represents a local gradient vector, 
i.e., $\nabla C  (\bw_{t-1}, \bx_{k})$.
For the quantization of $\bv$, the $l$th element of
$\bv$ can be expressed as
\be 
v_l = ||\bv|| \tilde v_l, 
\ee 
where 
$\tilde v_l = \frac{v_l}{||\bv||}$.
Suppose that $||\bv||$ and 
$\tilde v_l$ can be transmitted separately.
To encode $\tilde \bv = [\tilde v_1 \ \ldots \ \tilde v_L]^T$,
we can use a vector quantizer.
To this end, let
$\cC = \{\bc_1, \ldots, \bc_M \}$
be a codebook for vector quantization,
where $\bc_m$
represents the $m$th codeword.
Denote by ${\rm Conv} (\cC)$
the convex hull of the vectors in $\cC$,
i.e., ${\rm Conv}(\cC) = \{\sum_{m=1}^M a_m \bc_m \bigl|\ 
a_m \ge 0, \sum_{m}a_m = 1\}$.
Define the $L$-dimensional ball
of radius $R$ centered at $\bc$ as
$$ 
\cB_L(\bc, R)
= \{\bb\bigl| \ ||\bb - \bc||_2 \le R, \ \bb \in \uR^L\}. 
$$
Suppose that codebook $\cC$
satisfies the following condition:
\be
\cB_L (0, 1) \subseteq {\rm Conv}(\cC)
\subseteq \cB_L (0,R),
    \label{EQ:cond1}
\ee 
where $R > 1$. For a given vector 
$\tilde\bv \in \cB_L(0,1)$,
due to \eqref{EQ:cond1},
$\bv$ can be expressed by a convex
linear combination, i.e.,
\be
\tilde \bv = \sum_{m=1}^M a_m (\tilde \bv) \bc_m,
    \label{EQ:tbv} 
\ee 
where $a_m(\tilde \bv) \ge 0$ and 
$\sum_m a_m (\tilde \bv) = 1$.
Then,
the vector quantization
scheme in \cite{Gandikota19} is given by
\be 
Q_\cC (\tilde \bv) = \bc_m \ \mbox{w.p.} \ 
a_m(\tilde \bv),
	\label{EQ:Qm}
\ee 
where the 
convex combination weights,
the $a_m(\tilde \bv)$'s, are used as the probability distribution
to select a codeword
(i.e., $a_m$ is seen as the probability to choose $\bc_m$)
for given $\tilde \bv$.
The resulting quantizer is a randomized quantizer and
it can be readily shown that
the quantized vector is unbiased as
\be
\uE[Q_\cC (\tilde \bv)] = 
\sum_m a_m \bc_m = \tilde \bv,
\ee 
where $a_m = a_m (\tilde \bv)$.
Due to \eqref{EQ:cond1}, 
the MSE of $Q_\cC (\tilde \bv)$ is 
bounded as follows:
\begin{align}
\sMSE &  = \uE ||[\tilde \bv - Q_\cC (\tilde \bv) ||^2]    \cr
& = \sum_m ||\bc_m||^2 a_m  - ||\tilde \bv||^2 
\le R^2 - 1 \le R^2.
\end{align}
Once a device finds a codeword according to \eqref{EQ:Qm},
it sends the index of the codeword. Thus,
the number of bits 
to send $\tilde \bv$ is $\log_2 M$.

In \cite{Gandikota19}, it is shown that
a lower bound on $M$ grows exponentially with $L$ to meet
the condition for a fixed $R$ 
or $R^2$ grows linearly with
$L$ for a fixed $M$ in \eqref{EQ:cond1},
i.e.,
\be
M \ge \exp \left( c \frac{L}{R^2} \right),
	\label{EQ:lbM}
\ee
where $c > 0$ is constant.
Then, it can be shown that
\be
B (\sMSE)= \log_2 M \ge c^\prime \frac{L}{\sMSE},
\ee
where $c^\prime > 0$ is constant.
In Appendix A, we find bounds on $M$
for uniformly distributed codewords as follows:
\be
2 e^{\frac{L}{2 R^2}} \le M \le
\sqrt{2 \pi L} e^{\frac{L}{2 R^2}} \ \mbox{for a large $L$}.
	\label{EQ:bounds}
\ee
Thus, we can claim that 
$B$ is at most $O(L + \log \sqrt{L})$.
It is further shown that
if a Gaussian codebook is used for $\cC$,
the lower bound can be achieved. In this case,
when $\sMSE$ is a constant regardless of $L$,
we have $B= O(L)$ \cite{Gandikota19}.

As a deterministic construction for codebook, 
a scaled cross polytope (CP) is considered in \cite{Gandikota19}
as follows:
\be
\cC_{\rm cp} = \{ \pm R \bee_l: \ l \in \{1,\ldots, L\} \},
\ee
where $R = \sqrt{L}$ and $\bee_l$
represents the $l$th standard basis vector which 
has 1 in the $l$th position and 0 elsewhere.
Thus, $M = |\cC| = 2L$.

Note that since $||\bc_m|| = \sqrt{L}$ for all $\bc_m \in \cC_{\rm cp}$, 
the MSE is invariant with respect to 
the weights for convex linear combination
or the probabilities to select a codeword,
which is $\sMSE = L - 1$.
Throughout this paper, we assume 
that the CP codebook, $\cC_{\rm cp}$,
is used for quantization,
and let $\cC = \cC_{\rm cp}$ unless stated otherwise.

For convenience, 
the approach where each device in $\cK(t)$
transmits its encoded quantized vector with $B$ bits
through a dedicated sub-slot within a slot 
in a time division multiple access (TDMA) manner
is referred to as the conventional approach.
Then, the length or time of one iteration becomes
\be
T_{\rm conv} = \bar K  B
	\label{EQ:Tconv}
\ee
which shows that as mentioned earlier,  
the conventional approach can be slow  or inefficient
for a large minibatch size $\bar K$. 
It is noteworthy that in \eqref{EQ:Tconv},
the time to transmit the norms of gradient vectors is not included,
which is also proportional to $\bar K$.

\subsection{Quantization for Subvectors}	\label{SS:QS}

Although the CP codebook 
allows a closed-form expression for $\{a_m\}$,
it may not be suitable for the case that
the length of gradient vector, $L$, is large,
which results in a large MSE.
To decrease the MSE, as in 
\cite{Gandikota19}, the repetition can be used.
Alternatively, we can divide the gradient vector
into multiple subvectors and quantize each of them independently.


Let $\bv$ be divided into multiple sub-vectors as follows:
\be
\bv = [\bv_{(1)}^\rT \ \cdots \ \bv_{(D)}^\rT]^\rT,
\ee
where $\bv_{(d)} \in \uR^{\frac{L}{D}}$, where it is assumed
that $\bar L = \frac{L}{D}$ and $D$ are integers for convenience.
Then, each subvector can be quantized.
Let $\tilde \bv_{(d)} = \frac{\bv_{(d)}}{||\bv_{(d)}||}$.
The MSE of $Q_\cC (\tilde \bv_{(d)})$ becomes
$\bar L - 1$.
As a result, if the norms of the subvectors, $||\bv_{(d)}||$'s, 
are separately transmitted to the BS,
the MSE of the quantized gradient vector, denoted
by $\hat \bv$, becomes
\begin{align}
\sMSE (\hat \bv) & = \sum_{d = 1}^D ||\bv_{(d)}||^2
\sMSE ( Q_\cC (\tilde \bv_{(d)}) ) \cr
& = \sum_{d = 1}^D ||\bv_{(d)}||^2
(\bar L - 1) = 
||\bv||^2 \left( \frac{L}{D} - 1 \right).
	\label{EQ:Dcp}
\end{align}
On the other hand, the number of bits to encode $\tilde \bv$
becomes
\be
B = D \log_2 (2 \bar L) = D \log_2 \left( \frac{2 L}{D} \right).
	\label{EQ:Rcp}
\ee
From \eqref{EQ:Dcp} and \eqref{EQ:Rcp},
we can see a trade-off between the MSE and number of bits, $B$,
in the conventional approach.
That is, $D$ increases, the MSE decreases, while the number of 
bits increases.
It is noteworthy that 
the scheme that quantizes subvectors independently
provides lower MSE and smaller number of bits than the repetition
used in \cite{Gandikota19}. 
As will be shown later, this simple scheme
is also useful for the proposed random access based approach.

\section{Random Access for Parameter Updating over AWGN} \label{S:AWGN}

In this section, we propose an approach that
allows simultaneous transmissions to exploit the broadcast
nature of wireless communications so that
the time for one iteration is not 
necessarily proportional
to the number of devices, $K$.
As a result, the proposed 
approach is well-suited to the case of a large $K$.
Another salient feature is that there is no need
to quantize the norm of the gradient vector separately
(note that the quantization approaches in 
\cite{Alistarh17} \cite{Gandikota19}
need to separately send the norm).
Using the access probability in random access,
we can implicitly send the information of the norm of the gradient vector,
which makes the proposed approach communication-efficient.

We assume that the BS sends $\bw_{t-1}$ at the beginning of slot $t$
using downlink transmissions and $\cK(t) = \{1,\ldots, K\}$
(i.e., the minibatch size is the maximum, $\bar K= K$).
Then, each device
computes its gradient that is given by
$\bv_{k,t} = \nabla C (\bw_{t-1}, \bx_k)$,
and performs the quantization and send back 
the quantized gradient to the BS.
If all the gradient vectors can be received at the BS,
the next parameter vector becomes
\be
\bw_t = \bw_{t-1} - \mu \bg_t, 
	\label{EQ:AK}
\ee
where $\bg_t = \frac{1}{K}\sum_{k=1}^K \bv_{k,t}$.
Note that 
\eqref{EQ:AK}  is identical to \eqref{EQ:sgd}
with $\cK(t) = \{1,\ldots, K\}$.
Thus, the BS expects to have $\bg_t$
or its estimate,
which is an expensive option for the conventional approach
for a large $K$ in terms of communication cost.
In this section, based on the notion of random access,
we show that an estimate of $\bg_t$
becomes available in each upload regardless of $K$.
For convenience, we omit the time index $t$.

\subsection{Preamble-based Random Access}

In this section, we assume that the quantization
approach in Section~\ref{S:Quan} is used at devices
to quantize local gradient vectors.

Suppose that the BS knows the codeword
that is chosen by device $k$, which is denoted by $\bc_{m(k)} \in \cC$,
and the norm of the gradient, $||\bv_k||$, of each device.
Here, $m(k)$ represents the index of the codeword chosen by
device $k$.
If each device uploads $\{||\bv_k||, \bc_{m(k)}\}$
in a sequential manner, there should be $K$ uploads.
With $K$ uploads,
the BS can find the sum of quantized gradient vectors as follows:
\begin{align}
\hat \bg = \frac{1}{K} \sum_{k=1}^K  ||\bv_k|| \bc_{m(k)} 
= \frac{1}{K} \sum_{m=1}^M  \sum_{k \in \cK_m} ||\bv_k|| \bc_m,
	\label{EQ:dvq}
\end{align}
where $\cK_m = \{k\,:\, m(k) = m, \ k = 1, \ldots, K\}$.
Due to the randomized quantization, $m(k)$ is a random variable.
That is, according to \eqref{EQ:Qm}, we have
\be
m(k) = m \ \mbox{w.p. $a_m (\tilde \bv_k)$}.
\ee
From this, the mean of $\hat \bg$ is given by
\begin{align}
\uE[\hat \bg] 
& = \frac{1}{K} \sum_{k=1}^K ||\bv_k|| \uE[\bc_{m(k)}] \cr
& = \frac{1}{K} \sum_{k=1}^K ||\bv_k|| \sum_{m=1}^M \bc_m a_m (\tilde \bv_k) \cr
& = \frac{1}{K} \sum_{k=1}^K ||\bv_k|| \tilde \bv_k = \bg,
	\label{EQ:Ed}
\end{align}
which shows that $\hat \bg$ is an unbiased estimate
of the aggregation.

Note that according to \eqref{EQ:dvq},
the BS needs to know $\sum_{k \in \cK_m} ||\bv_k||$
in order to have $\hat \bg$.
To this end, based on the notion of random access, we propose  
a communication-efficient approach that does not need
$K$ separate uploads, but one (simultaneous) upload as follows.

Suppose that there are $M$ orthonormal preambles,
denoted by $\cP = \{\bp_1, \ldots, \bp_M\}$.
In addition, we assume that there is a one-to-one 
correspondence between the codebook, $\cC$, and
the preamble pool, $\cP$.
For convenience, it is assumed that 
if codeword $\bc_m$ is chosen, then a device
transmits preamble $m$.
In addition, each device
can decide whether or not it transmits depending
on the value of $||\bv_k||$.
To this end,
we assume that $||\bv_k|| \le V_{\rm max}$,
where $V_{\rm max}$
denotes the maximum norm of the gradient.
Then, let 
the access probability or the probability
that device $k$ transmits a preamble be
\be
p_k = \frac{||\bv||}{V_{\rm max}} \in [0,1].
	\label{EQ:pk}
\ee
In addition, define
\be
\beta_k = \left\{
\begin{array}{ll}
1, & \mbox{w.p. $p_k$} \cr
0, & \mbox{w.p. $1-p_k$.} \cr
\end{array}
\right.
	\label{EQ:nuk}
\ee
Then, the signal transmitted by device $k$ becomes
$\sqrt{P} \beta_k \bp_{m(k)}$,
where $P$ represents the transmit power.

The received signal at the BS over the
AWGN becomes
\be
\br = \sum_{k=1}^K \sqrt{P} \beta_k \bp_{m(k)} + \bn,
	\label{EQ:br}
\ee
where $\bn \sim \cN(0, \sigma^2 \bI)$ is the background noise.
Here, $m (k)$ becomes the index of preamble that is
chosen by device $k$ due to the one-to-one correspondence
between preambles in $\cP$ and codewords in $\cC$.
Since the preambles are orthonormal,
the output of the correlator becomes
\begin{align}
z_m & = \bp_m^\rT \br  
= \sqrt{P} \sum_{k=1}^K \beta_k \bp_m^\rT \bp_{m(k)} + \bp_m^\rT \bn \cr
& = \sqrt{P} \sum_{k=1}^K \beta_k \delta_{m, m(k)} + n_m 
= \sqrt{P} \sum_{k \in \cK_m} \beta_k + n_m, \ \
\end{align}
where $n_m =  \bp_m^\rT \bn \sim \cN(0, \sigma^2)$. From \eqref{EQ:pk}
and \eqref{EQ:nuk}, 
the conditional mean of $z_m$ is given by
\begin{align}
\uE[z_m \,|\, \cK_m] 
= \sqrt{P} \sum_{k \in \cK_m} \uE[\beta_k] 
= \frac{\sqrt{P}}{V_{\rm max} } \sum_{k \in \cK_m} ||\bv_k|| .
	\label{EQ:zKm}
\end{align}
Let
\begin{align}
\ba = \frac{1}{K} \sum_{m=1}^M z_m \bc_m.
	\label{EQ:ba1}
\end{align}
From \eqref{EQ:dvq} and \eqref{EQ:zKm}, the 
conditional mean of $\ba$ becomes
\begin{align}
\uE[\ba \,|\, \{\cK_m\}] & = 
\frac{1}{K}
\sum_{m=1}^M \uE[z_m \,|\, \cK_m]  \bc_m \cr
& = \frac{1}{K}
\frac{\sqrt{P}}{V_{\rm max}} \sum_{m=1}^M 
\sum_{k \in \cK_m} ||\bv_k|| \bc_m  
= \frac{\sqrt{P}}{V_{\rm max}} \hat \bg. \ \ 
\end{align}
Thus, using \eqref{EQ:Ed}, it can be
shown that the mean of $\ba$ is 
proportional to
the aggregation of the gradient vectors of $K$ devices as follows:
\begin{align}
\uE[\ba] = \uE[\uE[\ba \,|\, \{\cK_m\}]] 
= 
\frac{\sqrt{P}}{V_{\rm max}} \uE[ \hat \bg]
= 
\frac{\sqrt{P}}{V_{\rm max}} \bg.
\end{align}
From this, an unbiased estimate of $\bg$ can be obtained as
follows:
\be
\hat \bg = \frac{V_{\rm max}}{\sqrt{P}} \ba.
\ee
For convenience, the resulting approach will be referred to as 
the random access based updating scheme (RAUS).
The key feature of RAUS is to allow all devices
transmit their codewords simultaneously as random access
with the access probabilities that
are regarded as \emph{soft weights}.
Since the soft weights are linearly proportional to the norms of the gradient
vectors, 
the BS can have an estimate of $\bg$
without the norms explicitly transmitted by devices through different 
channels or time slots.

The length of slot in RAUS 
is equivalent to that of preambles. Since 
the preambles are orthogonal,
their length becomes the number of codewords in $\cC$, i.e., $M$. 
Thus, regardless of the number of devices, $K$,
the time for one iteration (that happen in one slot) in RAUS becomes
\be
T_{\rm raus} = O(M).
	\label{EQ:TRAUS}
\ee
With the cross polytope codebook, $\cC_{\rm cp}$,
from \eqref{EQ:Tconv} and \eqref{EQ:TRAUS}, we can show that 
\begin{align}
T_{\rm conv} & = O(K \log_2 (2L)) \cr
T_{\rm raus} & = O(2L).
\end{align}
Clearly, RAUS becomes more communication-efficient
than the conventional scheme when $K$ is large.
As a result, for comparison, 
we will not consider the conventional approach, but 
the approach in \cite{Yang20},
which will be discussed 
in Section~\ref{S:Fad}.

\subsection{RAUS with Multiple Preamble Transmissions}

As mentioned earlier, if $L$ is large,
the MSE of quantized gradient vector is large.
To avoid a large MSE, the gradient vector was divided
into $D$ subvectors in Subsection~\ref{SS:QS}.
In this case, multiple preambles are to be transmitted 
within
one round. The resulting approach is referred to as
the multiple preamble transmission (MPT) approach.
With the CP codebook,
we need a set of $\frac{2L}{D} = 2 \bar L$ (orthogonal) preambles
to transmit each sub-vector. Note that
since a total of $D$ sub-vectors are to be transmitted,
there are $D$ preambles of length $2 \bar L$ per one
gradient vector in MPT
with $\cC_{\rm cp}$,
which means that the time for one iteration 
is proportional to $2 \bar L D = 2 L$.
Clearly, unlike the conventional approach,
MPT does not reduce the 
time for one iteration in RAUS. 
However, as will be discussed later, MPT can 
reduce the steady-state MSE of RAUS.

%

\section{Random Access for Parameter Updating over
Fading Channels}	\label{S:Fad}

In this section, we discuss RAUS over fading channels.
Two different approaches are presented. The first approach
is based on the channel reciprocity. 
On the other hand, the second approach does not rely on
the channel reciprocity.

\subsection{Coherent Combining using CSI at Transmitter}

Let $h_k \in \uC$ denote the channel 
coefficient between the BS and device $k$.
Suppose that time division duplexing (TDD) mode is employed
so that the channel reciprocity can be exploited.
When the BS sends $\bw_{t-1}$ at the beginning of slot $t$,
suppose that it also transmits a downlink pilot signal so that
devices can estimate the channel coefficients.

Let $\phi_k = \frac{\sqrt{P} h_k^*}{|h_k|^2}$ be
the transmit gain at device $k$ for coherent combining 
at the BS.
Then, the received signal at the BS becomes
\begin{align}
\br = \sum_{k=1}^K h_k \phi_k \beta_k \bp_{m(k)} + \bn 
= \sum_{k=1}^K \sqrt{P} \beta_k \bp_{m(k)} + \bn,
\end{align}
which is identical to \eqref{EQ:br}.
This shows that RAUS can also be used for 
the system over fading channels.

Note that 
$|\phi_k|^2$ is the transmit power of device $k$.
Since the transmit power of mobile devices is limited,
we may have 
\be
|\phi_k|^2 \le P_{\rm max},
	\label{EQ:gk2}
\ee
where $P_{\rm max}$ represents the maximum transmit power of 
mobile devices. Thus, to take into account the 
transmit power constraint in \eqref{EQ:gk2}, 
$\beta_k$ can be modified as
\be
\beta_k =  
\left\{
\begin{array}{ll}
1, & \mbox{if ${\rm Unif}(0,1) \le p_k$ and $|\phi_k|^2  \le P_{\rm max}$} \cr
0, & \mbox{o.w.} \cr
\end{array}
\right.
\ee
Here, ${\rm Unif}(0,1)$ represents a uniform 
random variable between 0 and 1.

\subsection{Noncoherent Combining with Multiple Antennas}	\label{SS:NC}

Suppose that the channel reciprocity cannot be exploited. 
In this case, 
devices are unable to decide their transmit gains
for coherent combining at the BS.
Thus, device $k$ transmits $\beta_k \bp_{m(k)}$
as in Section~\ref{S:AWGN}.

We assume that the BS is equipped with
multiple antennas. Let $N$ 
represent the number of antennas and
denote by $\bh_k \in \uC^{N}$ 
the channel from the $k$th device to the BS.
Then, the received signal at the BS is given by
\be
\bR = \sum_{k=1}^K \sqrt{P_k} \beta_k \bh_k \bp_{m(k)}^\rT + \bN
\in \uC^{N \times L},
\ee
where $P_k$ represents the transmit power of device $k$
and $[\bN]_{n,l} \sim \cC \cN(0, N_0)$ is the background noise. 
The output of the correlator
with $\bp_m$ becomes
\begin{align}
\bz_m = \bR \bp_m 
= \sum_{k\in \cK_m} \sqrt{P_k} \beta_k \bh_k + \bn_m,
\end{align}
where $\bn_m = \bN \bp_m \sim \cC \cN(0, N_0 \bI)$.

As in \cite{Marzetta10}, suppose that
$\bh_k \sim \cC \cN (\b0, \alpha_k \bI)$,
where $\alpha_k$ is the large-scale fading term
that depends on the distance between the BS and device $k$.
If $P_k$ is decided to compensate the large-scale fading term,
i.e., $P_k \alpha_k = P$ for all $k$,
for given $\cK_m$, we have
\be
\bz_m \sim  \cC \cN (\b0, \bSigma_m),
	\label{EQ:zN}
\ee
where 
\be
\bSigma_m =\left( P \sum_{k \in \cK_m} 
\beta_k + N_0 \right) \bI.
\ee
For noncoherent combining, we can use 
$||\bz_m||^2$ that has the following conditional mean:
\be
\frac{\uE[||\bz_m||^2 \,|\, \cK_m]}{N} =
P \sum_{k \in \cK_m} \beta_k + N_0.
	\label{EQ:slln}
\ee
To obtain an estimate of the aggregation
using noncoherent combining, let
\be
\ba = \frac{1}{K} \sum_{m=1}^M \frac{||\bz_m||^2}{N} \bc_m.
	\label{EQ:ba2}
\ee
From \eqref{EQ:slln},
since $\uE[\beta_k] = \frac{||\bv_k||}{V_{\rm max}}$,
it can be readily shown that
\begin{align}
\uE[\ba \,|\, \{\cK_m\}] 
& = \frac{P}{K V_{\rm max}} \sum_{m=1}^M
\sum_{k \in \cK_m} ||\bv_k|| \bc_m
+ \sum_{m=1}^M \frac{N_0}{K} \bc_m \cr
& = \frac{P}{V_{\rm max}} \hat \bg + \frac{N_0}{K} \sum_{m=1}^M \bc_m.
\end{align}
If the sum of codewords is zero
(which is the case of $\cC = \cC_{\rm cp}$), the mean of $\ba$ 
becomes 
\be
\uE[\ba] = \frac{P}{V_{\rm max} K}
\sum_{k=1}^K \bv_k =  \frac{P}{V_{\rm max}} \bg.
\ee
As a result, we can have an unbiased estimate of the aggregation
as follows:
\be
\hat \bg = \frac{V_{\rm max}}{P} \ba.
	\label{EQ:hg2}
\ee

For the convergence analysis, suppose that
\be
\kappa \bI 
\preceq 
\nabla^2 f(\bw) 
\preceq U \bI,
	\label{EQ:bHess}
\ee
where 
$\kappa, U > 0$,
i.e., $f(\bw)
= \frac{1}{K} \sum_{k=1}^K
C(\bw; \bx_k)$ is $\kappa$-strongly and $U$-smooth convex.
Define the MSE of $\hat \bg$ as
\be
\sMSE (\hat \bg)
= \uE[||\hat \bg-\bg||^2]
= \uE[||\hat \bg||^2] - ||\bg||^2,
\ee
where the second
equality is valid if $\hat \bg$ is an unbiased estimate of $\bg$.
If $\sMSE (\hat \bg)  \le \bar \sigma_\bg^2 < \infty$,
it can be shown that
\be
\uE[||\bw_t - \bw^*||^2]
\le \frac{\mu \bar \sigma_\bg^2}{ (2 -  \mu \kappa) \kappa }.
	\label{EQ:ball}
\ee
In Appendix B, with the cross 
polytope codebook, $\cC_{\rm cp}$,
we can show that 
\be
\uE[||\hat \bg||^2 ] 
\le 
L \left(1 + \frac{2}{N} \right)
\left( U + \frac{P M N_0}{ V_{\rm max} K} \right)^2.
	\label{EQ:BB}
\ee
As a result, the parameter vector will converge to a noise ball
as in \eqref{EQ:ball}.
However, from \eqref{EQ:BB}, we do not see that 
$\uE[||\hat \bg - \bg||^2]$ decreases with the size of minibatch, $K$.
Thus, 
we consider the asymptotic case such as massive 
multiple-input multiple-output
(MIMO) \cite{Marzetta10}
where $N \to \infty$ to gain insight into RAUS.



For a large $N$ (i.e., massive MIMO), 
we can see that
\be
\frac{||\bz_m||^2}{N} \to P \sum_{k \in \cK_m} \beta_k + N_0,
\ N \to \infty.
\ee
Then, 
from \eqref{EQ:ba2}
and \eqref{EQ:hg2},
the asymptotic aggregation $\hat \bg$ can be given by
\be
\hat \bg = \frac{V_{\rm max}}{P} \ba 
\to 
\left(\frac{V_{\rm max}}{K} \sum_{k=1}^K
\beta_k \bc_{m(k)}
\right), \ N \to \infty.
	\label{EQ:Ahg}
\ee
As derived in Appendix C, we can show that
\begin{align}
\uE\left[|| \hat \bg - \bg||^2 \right]
& = \frac{1}{K^2}
\sum_{k=1}^K \left( L V_{\rm max} - ||\bv_k||\right) 
||\bv_k|| \cr
& \le \frac{L V_{\rm max}^2}{K},
	\label{EQ:AMSE}
\end{align}
thanks to $||\bv_k|| \le V_{\rm max}$.
From \eqref{EQ:AMSE},
we can see that
RAUS can effectively have the maximum size of minibatch,
$K$, without any additional communication cost
(i.e., the time for one round is fixed regardless of $K$) 
by exploiting the notion of random access.
As a result, RAUS can avoid the dilemma stated in
Section~\ref{S:SM}.

Note that in \eqref{EQ:AMSE}, 
if MPT is used, 
$L$ is replaced with
$\bar L$ and $V_{\rm max}$ is also replaced with
its counterpart, denoted by $\bar V_{\rm max}$, i. e.,
$||\bv_{(d)} || \le \bar V_{\rm max}$.
If $\bar V_{\rm max} = \frac{V_{\rm max}}{\sqrt{D}}$,
we can see that the MSE decreases by a factor of $1/D$.
Thus, MPT can effectively reduce the MSE of $\hat \bg$
without increasing the time for one iteration.

\section{Comparison with An Over-the-Air Computation Approach}	\label{S:Yang}

In this section, for comparison with RAUS in 
Subsection~\ref{SS:NC},
we consider an approach that also assumes a large number of 
antennas at the BS.

The approach in \cite{Yang20} 
based on the notion of 
over-the-air computation 
allows simultaneous transmissions 
by multiple devices in one iteration.
However, this approach requires
the estimation of the CSI of devices at the BS
using uplink pilot signals
transmitted by devices
and the feedback to the devices.


While the minimization
of the MSE of the estimate of the aggregation based
a non-convex optimization
formulation is studied in \cite{Yang20},
we consider a simplified version for comparison.
Let $\bm \in \uC^N$ denote the beamforming vector to combine
the signals from devices. Recall that $\cK (t)$
is a random subset of $\{1, \ldots, K \}$.
For convenience,
we omit the time index $t$.
Consider
the received signal at the BS 
when one of the elements of $\bv_k$, denoted by
$v_{k;l}$, is transmitted, 
which is given by
\be
\br_l = \sum_{k \in \cK} \sqrt{P_k} \bh_k \psi_k v_{k;l} + \bn_l,
\ee
where $\psi_k \in \uC$ is 
the phase compensation coefficient of device $k$ 
for coherent combining with $|\psi_k| = 1$.
Then, assuming that 
$||\bm|| = 1$, the (scaled) estimate of $g_l$, denoted by
$a_l$, becomes
\begin{align}
a_l & = \Re(\bm^\rH \br_l) \cr
& = \sum_{k \in \cK} \sqrt{P_k} \Re(\bm^\rH \bh_k \psi_k) v_{k;l} + n_l,
	\label{EQ:Yang_r}
\end{align}
where $n_l = \Re(\bm^\rH \bn_l) \sim \cN(0, N_0/2)$.
According to \eqref{EQ:Yang_r},
for coherent combining,
it is desirable that
$P_k |\bm^\rH \bh_k|^2 = C$ for all $k$, or
\begin{align}
\sqrt{P_k} \bm^\rH \bh_k \psi_k  = \sqrt{C},
	\label{EQ:Yang_Req}
\end{align}
i.e., $\psi_k = e^{-j \theta_k}$, where $\theta_k = \angle (\bm^\rH \bh_k)$.
With the assumption that $\bar K \ll N$
(recall that $\bar K = |\cK|$ is the size of minibatch),
it is known that the $\bh_k$'s 
are asymptotically orthogonal
to each other as in \cite{Marzetta10}.
Thus, 
assuming that all the channel vectors are orthogonal,
to satisfy \eqref{EQ:Yang_Req}
with a maximum of $C$,
$\bm$ can be found as
\be
\hat \bm = \frac{\sum_{k \in \cK} \sqrt{P_k} \bh_k}{||
\sum_{k \in \cK} \sqrt{P_k} \bh_k||}.
\ee
With $\sqrt{P_k} \bh_k \sim \cC \cN (\b0, P \bI)$, where
$P = P_k \alpha_k$ for all $k$ (as assumed earlier in
Subsection~\ref{SS:NC}), 
we can show that
\be
P_k |\bm^\rH \bh_k|^2 \to \frac{P N}{ \bar K},
 N \to \infty.
\ee
As a result, for comparison,
we will consider the following 
asymptotic approximation of
\eqref{EQ:Yang_r}:
\begin{align}
\ba & = [a_1 \ \ldots \ a_L]^\rT 
= \sqrt{\frac{P N}{\bar K} } 
\sum_{k \in \cK} \bv_k + \bn \cr
& = \sqrt{ P N \bar K}  
\left( \frac{1}{\bar K} \sum_{k \in \cK} \bv_k \right) + \bn,
\end{align}
where $\bn = [n_1 \ \ldots \ n_L]^\rT$.
Letting $\hat \bg = \frac{1}{\sqrt{PN \bar K}} \ba$ as
the estimate of the aggregation,
we can show that 
\begin{align}
\uE[||\hat \bg - \bg||^2] 
& =  \frac{ \uE[||\bn||^2] }{PN \bar K} + \uE\left[||\bg - 
\frac{1}{\bar K} \sum_{k \in \cK} \bv_k||^2 \right] \cr
& = \frac{1}{\bar K \frac{2 P}{N_0} }
+ \frac{1}{\bar K} \uE_k [||\bg - \bv_k||^2] \cr
& = \frac{1}{\bar K}
\left(\frac{N_0}{2 P}
+ \left(\frac{1}{K} \sum_{k=1}^K ||\bv_k||^2 - ||\bg||^2
\right) \right), \quad
	\label{EQ:MSEota}
\end{align}
where $\uE_k [\cdot]$ represents the expectation over the index
$k$ that is uniformly distributed over $\{1,\ldots, K\}$.
This shows that the MSE is $O(1/\bar K)$.

As mentioned earlier,
in order to find $\bm$ satisfying \eqref{EQ:Yang_Req}, 
\emph{i)}
the CSI of $\bar K =|\cK|$ devices per each
round at the BS should be known
(which requires uplink pilot transmissions from the
devices in $\cK$);
\emph{ii)} the feedback of $\psi_k$ to the devices belonging
to $\cK$ in each round is required.

To see the time for one round or iteration,
let $\tau_{\rm pilot}$ be the 
length of uplink pilot from each device. Then, the total
time of pilot transmissions of $\bar K$ devices becomes
$\bar K \tau_{\rm pilot}$.
As a result, 
the time for one iteration of
the approach in \cite{Yang20} 
becomes
\be
T_{\rm ota} = \bar K \tau_{\rm pilot} + c L,
	\label{EQ:Tota}
\ee
where $c > 0$ is constant.
Here, $cL$ represents the time to transmit local gradient
vectors by devices, which is linearly proportional to the length
of $\bv$, $L$, as shown in \eqref{EQ:Yang_r}.
Thus, as $\bar K = |\cK (t)|$ increases,
the time for one iteration increases.
Note that in \eqref{EQ:Tota}, we do not include the time
for feedback (of $\psi_k$, $k \in \cK$) from the BS to devices,
meaning that \eqref{EQ:Tota} can be seen as a lower-bound.

From \eqref{EQ:MSEota} and 
\eqref{EQ:Tota}, we can observe that
the approach in \cite{Yang20}
cannot overcome the dilemma mentioned in Section~\ref{S:SM},
i.e., $\bar K$ cannot be increased or decreased,
although the notion of over-the-air computation is exploited.
On the other hand, as mentioned earlier, RAUS 
does not have this problem
and the size of minibatch can be the maximum, i.e., $\bar K = K$.
This means that
all the devices can participate in the upload and send local gradients
simultaneously,
without increasing the time for one iteration.

\section{Simulation Results}	\label{S:Sim}

In this section, we present simulation results
under various conditions.
For comparison with
the approach in \cite{Yang20},
we only consider noncoherent combining 
in Subsection~\ref{SS:NC} for RAUS
with a BS 
equipped with multiple antennas.
In addition, it is assumed that $\sqrt{P_k} \bh_k
\sim \cC \cN (\b0, P\bI)$ for all $k$ as in
Subsection~\ref{SS:NC}.

The MSE of the estimate of the aggregation,
i.e., $\uE[||\hat \bg - \bg||^2]$, is used
as a performance metric.
For convenience, the approach
in \cite{Yang20} is referred to as YANG.
The signal-to-noise ratio
(SNR) is defined as $\frac{P}{N_0}$.

\begin{figure}[thb]
\begin{center}
\includegraphics[width=\figwidth]{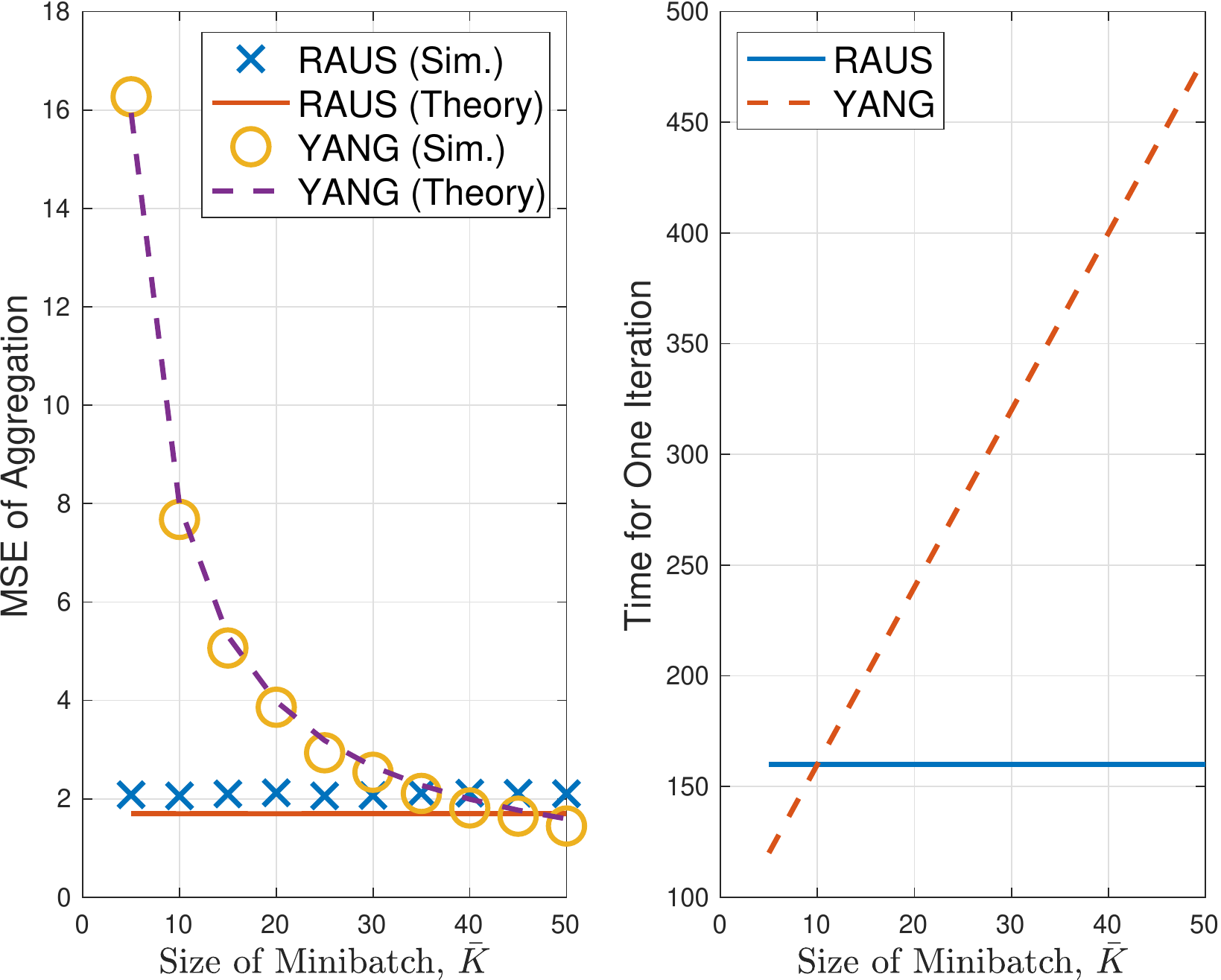} \\
\hskip 0.5cm (a) \hskip 3.5cm (b)
\end{center}
\caption{Performance of the RAUS and YANG approaches
for different size of minibatch
with $L = 80$ ($\bar L = 8$ and $D = 10$), $K = 500$, $N = 100$,
and SNR = 4 dB: 
(a) MSE; (b) Time for one iteration.}
        \label{Fig:plt_sim2}
\end{figure}

In Fig.~\ref{Fig:plt_sim2} (a),
the MSE of $\hat \bg$ is shown 
as a function of the size of minibatch, $\bar K$,
where $L = 80$, $K = 500$, and $N = 100$, SNR $= 4$ dB.
For RAUS, MPT is considered with
$\bar L = 8$ and $D = 10$. 
The theoretical MSEs of the RAUS and YANG approaches
are given by \eqref{EQ:AMSE} and \eqref{EQ:MSEota}, respectively.
Since RAUS is independent of $\bar K$ (the effective size
of minibatch in RAUS is $K$),
the MSE of $\hat \bg$ in RAUS
is constant, while
that in YANG decreases with $\bar K$ as expected.
However, 
as shown Fig.~\ref{Fig:plt_sim2} (b),
the time for one iteration, $T_{\rm ota}$, in YANG increases.
For $T_{\rm ota}$ in \eqref{EQ:Tota},
we assume that $\tau_{\rm pilot}$
is 10\% of the time to transmit one gradient vector,
i.e., $T_{\rm ota} = (0.1 \bar K + 1)L$,
with $c = 1$.
On the other hand, the time for one iteration in RAUS
is set to $T_{\rm raus} = 2 L$,
which is independent of the minibatch size.
Clearly, we can see that RAUS can have a smaller 
MSE than YANG with a constant time for one iteration
regardless of the total number of devices, $K$.

\begin{figure}[thb]
\begin{center}
\includegraphics[width=\figwidth]{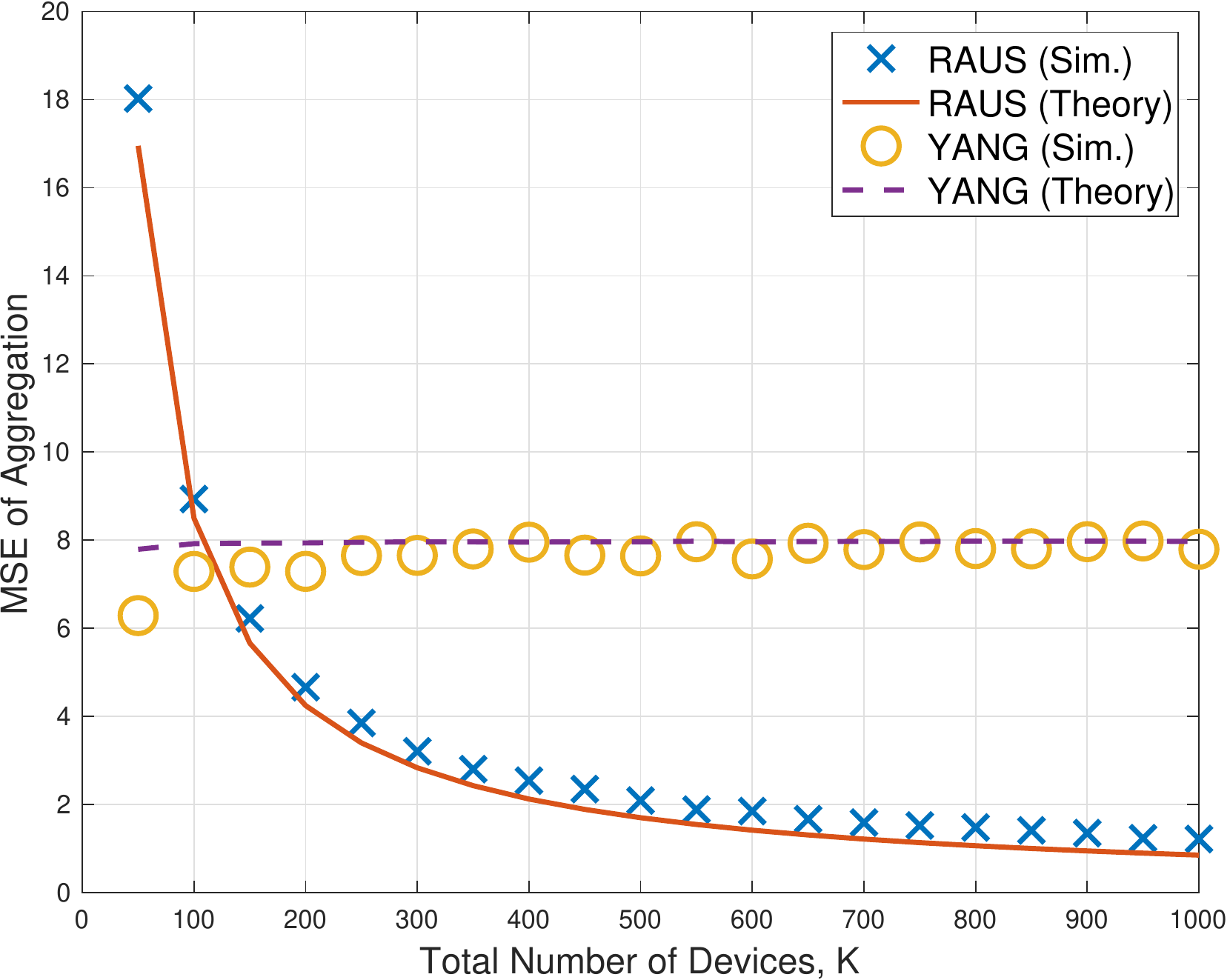} 
\end{center}
\caption{Performance of the RAUS and YANG approaches
for different number of devices, $K$
with $L = 80$ ($\bar L = 8$ and $D = 10$), $\bar K = 10$, $N = 100$,
and SNR = 4 dB.} 
        \label{Fig:plt_sim1}
\end{figure}

Fig.~\ref{Fig:plt_sim1}
shows the impact of the total number of devices, $K$,
on the MSE in RAUS and YANG
with $L = 80$ ($\bar L = 8$ and $D = 10$ for RAUS),
$\bar K = 10$, and $N = 100$. 
Note that the time for one iteration in YANG
depends on the size of minibatch, which is set to $\bar K = 10$.
As a result, the time for one iteration in both the approaches 
is constant regardless of $K$.
We see that the MSE in RAUS decreases with $K$ as expected.
Clearly, it shows that RAUS is a communication-efficient approach
for distributed SGD when the number of devices, $K$, is large.

\begin{figure}[thb]
\begin{center}
\includegraphics[width=\figwidth]{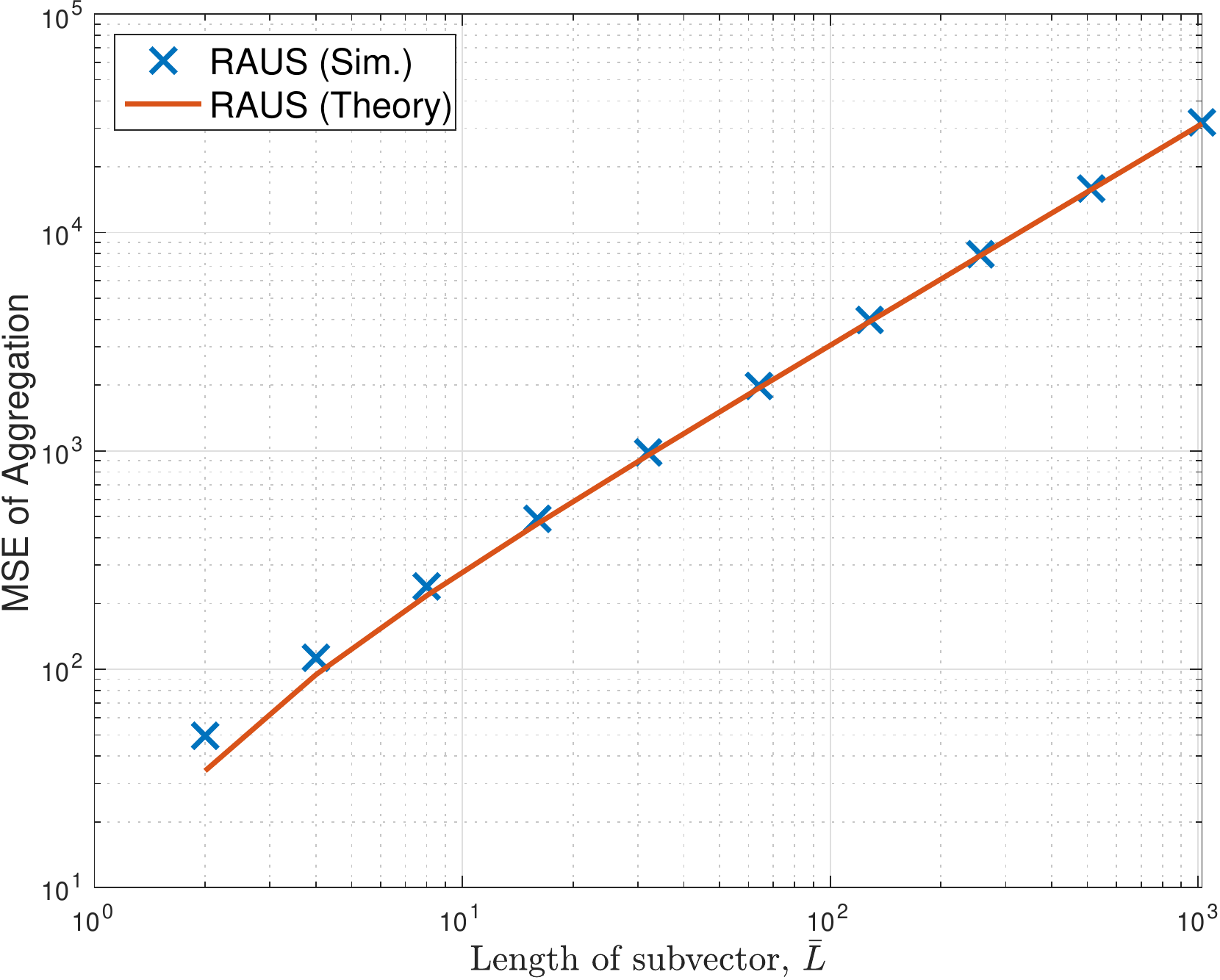} 
\end{center}
\caption{Performance of RAUS 
as a function of $\bar L$ when $L$ is fixed
with $L = 2^{12} = 4096$, $K = 200$, $N = 100$, and 
SNR = 4 dB.}
        \label{Fig:plt_sim3}
\end{figure}

As explained in Subsection~\ref{SS:NC},
MPT can reduce the MSE in RAUS.
To see this, with a fixed $L = 4096$,
the MSE is obtained with increasing $\bar L$.
The results are shown in Fig.~\ref{Fig:plt_sim3}.
Clearly, in order to decrease the 
MSE, we need to keep $\bar L$ small.
This is also useful for devices with limited
storage as the size of
codebook or preamble pool increases with $\bar L$.

To see the performance of RAUS and YANG in distributed
SGD, we consider the support vector classifier (SVC)
with training image data sets in \cite{Krizhevsky09}.
Only two different classes of image data sets
are considered for binary linear SVC. The length
of the parameter vector is $L = 32 \times 32 \times 3 = 
3072$\footnote{Note that the length of gradient vector is actually
$L = 3072+1$ due to the offset term. In RAUS, a codebook of 2 elements
is considered to send the coefficient corresponding to the offset term.}
(the size of image is $32 \times 32$ and each image has 3 different
colors).
As in \cite{Bishop06},
the cost function based on the hinge function is given by
\be
C(\bw, \bx_k) =  \max(0, 1 - \ell_k (\bw^\rT \bx_k-w_0)) + \lambda ||\bw||^2,
\ee
where $\ell_k \in \{\pm 1\}$ is the label of data set at device $k$
(i.e. $\ell_k = -1$ and $+1$ for labels 0 and 1, respectively),
$w_0$ represents the offset term,
and $\lambda > 0$ is the Lagrange multiplier.
We assume that each device has one image
and there are $K/2 = 500$ devices with label 0 and
$K/2 = 500$ with label 1, i.e., there are a total of $K = 1000$
devices.
In addition, for simulations, we assume that $N =100$ and SNR $=10$ dB.

\begin{figure}[thb]
\begin{center}
\includegraphics[width=\figwidth]{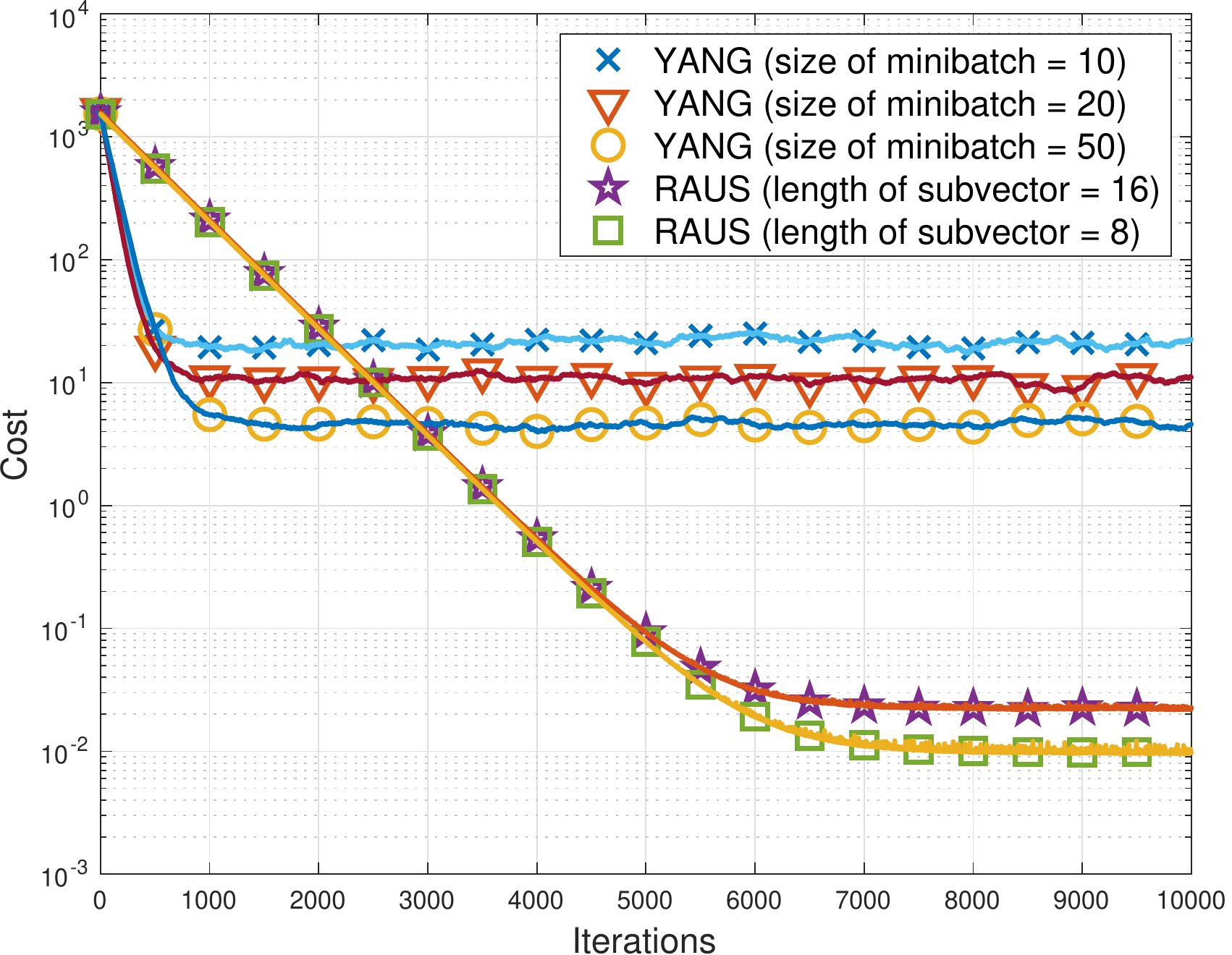} 
\end{center}
\caption{Convergence of distributed SGD for binary SVC
when YANG and RAUS are used to upload devices' local gradient vectors
with $K = 1000$, $N = 100$, and SNR $= 10$ dB.}
        \label{Fig:plt_svm2}
\end{figure}

In Fig.~\ref{Fig:plt_svm2},
it is shown that the cost decreases as the number of rounds increases
(up to $T = 10,000$).
For YANG and RAUS, the step-size is
set to $\mu = 0.01$ and $0.1$, respectively.
Note that the step-size in YANG is smaller than that in RAUS,
because the size of minibatch, $\bar K$, is usually smaller
than the total number of devices, $K$. In particular,
we consider $\bar K \in \{10,20, 50\}$ for YANG.
Clearly, as shown in
Fig.~\ref{Fig:plt_svm2},
the steady-state cost decreases with $\bar K$ in YANG,
which is expected from Fig.~\ref{Fig:plt_sim2}.
For RAUS, we have two different values of $\bar L$, i.e.,
$\bar L \in \{8,16\}$.
Since the MSE of the estimated aggregation
decreases as $\bar L$ decreases
as shown in Fig.~\ref{Fig:plt_sim3},
we can see that the steady-state cost becomes smaller as
$\bar L$ decreases in RAUS.

\section{Concluding Remarks}	\label{S:Con}

For communication-efficient distributed SGD
over wireless channels,
we proposed an approach based on random access.
In particular, in the proposed approach
with a preamble-based random access scheme,
we considered a one-to-one correspondence between 
the quantization codebook, which is used 
for quantizing local gradient vectors,
and the preamble set, which is used for random access,
so that a device can send a preamble corresponding
to its quantized gradient vector.
In addition, as soft weights, the access probability has been controlled
to send the information of the norm of gradient vector
implicitly without using additional channel resources.

We showed that the proposed approach 
can support a large number of devices
participated in distributed SGD without increasing
the time for one iteration.
In fact, the performance can be improved by increasing
the number of devices as the MSE of the estimated aggregation 
decreases with the number of devices.
The MSE of the estimated aggregation was also analyzed
and compared with that in \cite{Yang20}. From simulations,
we also confirmed that the theoretical MSE obtained
by asymptotic analysis is close 
to simulation results.

\appendices

\section*{Appendix A: Bounds on $M$}

In \cite{Gandikota19}, it
is shown that $\cB_L (0, 1) \subseteq {\rm Conv}(\cC)$
if and only if
there exists $\bc \in \cC$ such that
$\langle \bx, \bc \rangle \ge 1$ for any $\bx \in \cS^{L-1}$.
Here, $\cS^n$ represents the $n$-sphere, i.e.,
$\cS^n = \{\bx \in \uR^{n+1}: \ ||\bx || = 1\}$.
Thus, for uniformly distributed codewords, $\bc_m$'s, 
\be
|\cC| = \frac{\mbox{Area ($\cS^{L-1}$)}}{\mbox{Area ($\Delta$)}},
\ee
where $\Delta = \{\bx \in \cS^{L-1}: \ \langle \bx, \bc \rangle \ge 1\}$.
From \eqref{EQ:cond1},
we have $||\bc|| \le R$. Thus, 
$\Delta$ is the (hyper-spherical) cap
with angle $\phi$ such that $\cos \phi \ge
\frac{1}{||\bc||} \ge \frac{1}{R}$.

\begin{figure}[thb]
\begin{center}
\includegraphics[width=\figwidth]{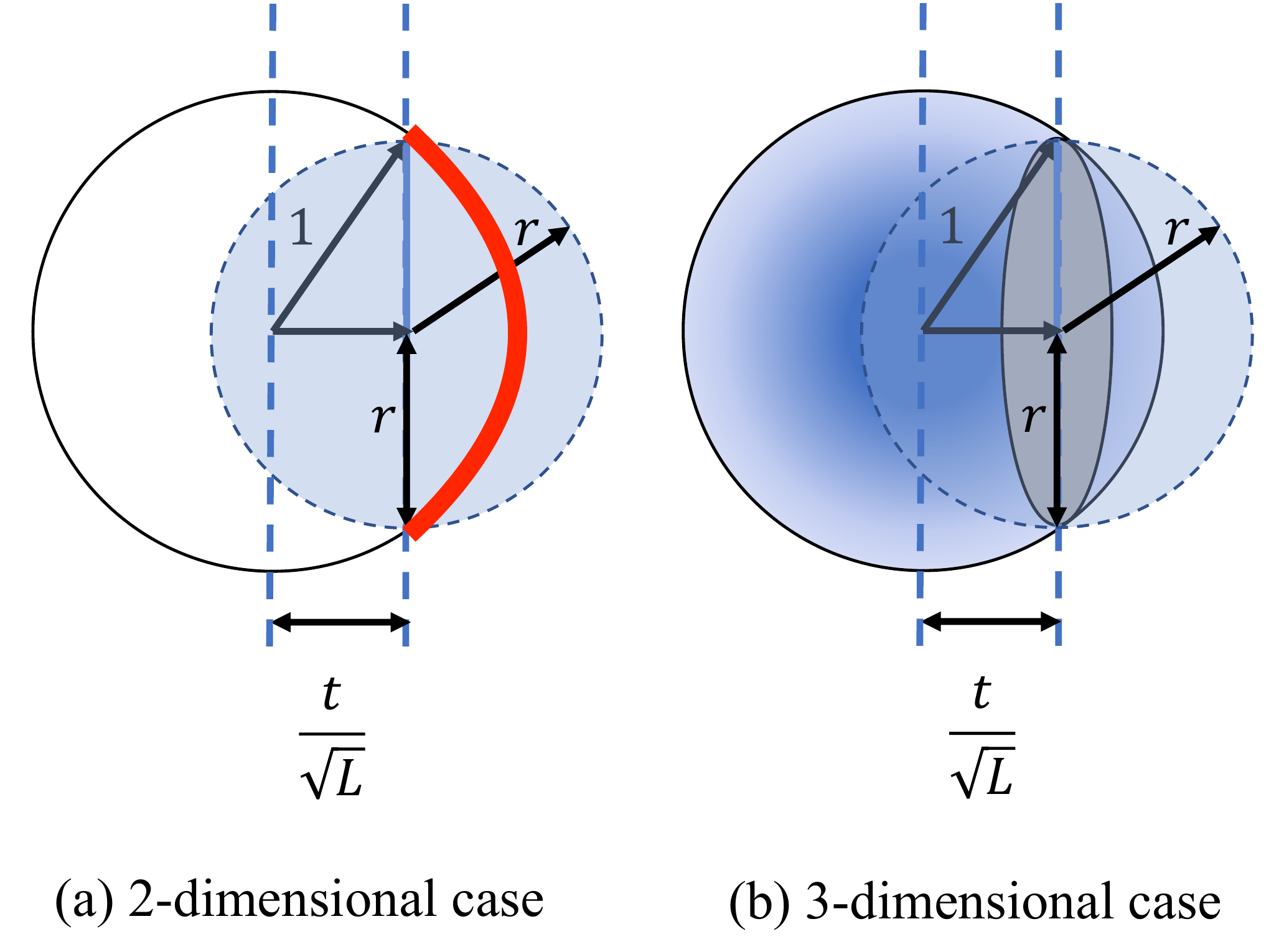}
\end{center}
\caption{Upper and lower bounds on the area of cap:
(a) the case of $L = 2$;
(b) the case of $L = 3$.}
        \label{Fig:caps}
\end{figure}

In Fig.~\ref{Fig:caps} (a) and (b), 
the areas of the cap are shown for the cases of $L = 2$ 
(by the thick line)
and $L = 3$, respectively, with $\frac{1}{R} = \frac{t}{\sqrt{L}}$.  
The area of the cap is bounded by
a half surface of a sphere of radius $r$
(upper-bound)
and the volume of the slice of radius $r$.
For the case of $L = 2$, 
we have
$2r < \mbox{Area of cap} < \pi r$,
where $2r$ and $\pi r$ are the lower and upper bounds, respectively.
Furthermore, for $L = 3$, 
$ \pi r^2 < \mbox{Area of cap} < 2 \pi r^2$.
For any $L$, it can be shown that
\begin{align}
V_{L-1} (r) 
= \frac{\pi^{\frac{L-1}{2}}}{\Gamma\left(
\frac{L+1}{2} \right)} r^{L-1}
& < \mbox{Area of cap} \cr
&  < \frac{S_L (r)}{2}
= \frac{\pi^{\frac{L}{2}} }{\Gamma \left( \frac{L}{2} \right)} r^{L-1}.
\end{align}

Let $\rho = \frac{1}{|\cC|}$. Then, we have
\begin{align}
\rho
& \le \rho_{\rm ub} = \frac{S_L (r)}{2 S_L (1)} = \frac{1}{2}
r^{L-1} = \frac{1}{2} 
\left(1 - \frac{t^2}{L} \right)^{\frac{L-1}{2}} \cr
\rho  
& \ge \rho_{\rm lb} = \frac{V_{L-1} (r)}{S_L (1)} =
\frac{\Gamma \left(\frac{L}{2} \right)}{
\Gamma \left( \frac{L+1}{2} \right) 2 \sqrt{\pi}}
\left(1 - \frac{t^2}{L} \right)^{\frac{L-1}{2}}.
\end{align}
With $t = \frac{\sqrt{L}}{R}$, it can be shown that
\be
2 \left(1 - \frac{t^2}{L} \right)^{-\frac{L-1}{2}}
\le \frac{1}{\rho} \le
\sqrt{2 \pi L} \left(1 - \frac{t^2}{L} \right)^{-\frac{L-1}{2}}.
\ee
or, for a large $L$,
\be
2 e^{\frac{L}{2 R^2}} \le |\cC| \le
\sqrt{2 \pi L} e^{\frac{L}{2 R^2}}.
\ee

\section*{Appendix B: MSE of $\hat \bg$ in \eqref{EQ:hg2}}

Since $\hat \bg$ in \eqref{EQ:hg2}
is unbiased,
we have
\begin{align}
\sMSE (\hat \bg) 
& = \uE\left[ ||\hat \bg - \bg||^2 \right] 
= \uE[||\hat \bg||^2] - ||\bg||^2 \cr
& = \left( \frac{V_{\rm max}}{P} \right)^2
\left( \uE[||\ba||^2] - ||\uE[\ba]||^2 \right).
	\label{EQ:a_vg}
\end{align}
Thus, to show that 
the variance of $\hat \bg$ is finite,
it is sufficient to find $\uE[||\ba||^2]$.

From \eqref{EQ:ba2}, it can be shown that
\begin{align}
\uE[||\ba||^2]
& = \uE \left[ ||\sum_m \frac{||\bz_m||^2 }{KN} \bc_m||^2 \right] \cr
& = \frac{1}{K^2}
\biggl(
\sum_m \frac{\uE[||\bz_m||^4]}{N^2} ||\bc_m||^2  \cr
& \ \ + 
\sum_{m \ne m^\prime} \frac{\uE[||\bz_m||^2]\uE[||\bz_{m^\prime}||^2] }{N^2} 
\bc_m^\rT \bc_{m^\prime} 
\biggl).
\end{align}
From \eqref{EQ:zN},
we have
\begin{align}
\frac{\uE[||\bz_m||^4]}{N^2} & = 3 \sigma_m^4 \cr
\frac{\uE[||\bz_m||^2]\uE[||\bz_{m^\prime}||^2] }{N^2} & = 
\sigma_m^2 \sigma_{m^\prime}^2,
\end{align}
where $\sigma_m^2 = \frac{P}{V_{\rm max}} \sum_{k \in \cK_m} ||\bv_m|| + N_0$.
Then, after some manipulations,
it can be shown that
\be
K^2 \uE[||\ba||^2] =
||\sum_m \sigma_m^2 \bc_m ||^2 + \frac{2}{N} 
\sum_m \sigma_m^4 ||\bc_m||^2.
	\label{EQ:a2_4}
\ee
Since $\bc_m \in \cC_{\rm cp}$,
we have $||\bc_m||^2 = L$ and
\begin{align}
||\sum_m \sigma_m^2 \bc_m ||^2 
& = \sum_{l=1}^L (\sigma_l^2 - \sigma_{L+l}^2 )^2 L \cr
& \le \sum_{l=1}^L (\sigma_l^4 + \sigma_{L+l}^4 ) L 
= L \sum_m \sigma_m^4 .
	\label{EQ:a2_5}
\end{align}
Substituting \eqref{EQ:a2_5} into \eqref{EQ:a2_4},
we have
\be
\uE[||\ba||^2] 
\le \frac{L \left(1+ \frac{2}{N} \right)}{K^2} 
\sum_{m} \sigma_m^4.
\ee

With $\sum_m \sigma_m^2 = c_1$,
it can be shown that
\begin{align}
\sum_m \sigma_m^4
&  \le  c_1^2 = 
\left(\sum_m 
 \frac{P}{V_{\rm max}} \sum_{k \in \cK_m} ||\bv_m|| + N_0
\right)^2 \cr
& = \left( \frac{P}{V_{\rm max}}\sum_k ||\bv_k|| + M N_0 \right)^2.
\end{align}
From \eqref{EQ:bHess},
we have $||\bv_k || = ||\nabla C (\bw, \bx_k) || \le U$.
Then, it follows
that
\begin{align}
\uE[||\ba||^2] 
& \le L \left(1 + \frac{2}{N} \right)
 \left( \frac{\frac{P}{V_{\rm max}}
\sum_k ||\bv_k|| + M N_0}{K}\right)^2 \cr
& \le
L \left(1 + \frac{2}{N} \right)
\left( \frac{P}{V_{\rm max}} U
+ \frac{M N_0}{K} \right)^2.
	\label{EQ:Ea2}
\end{align}
Substituting \eqref{EQ:Ea2} into \eqref{EQ:a_vg},
we have \eqref{EQ:BB}.

\section*{Appendix C: Derivation of \eqref{EQ:AMSE}}

From \eqref{EQ:Ahg}, 
we have
\begin{align}
\hat \bg - \bg = 
\frac{1}{K} \sum_{k=1}^K
\left( V_{\rm max} \beta_k \bc_{m(k)} - ||\bv_k|| \tilde \bv_k \right).
\end{align}
Since
$$
\uE\left[ V_{\rm max} \beta_k \bc_{m(k)} - ||\bv_k|| \tilde \bv_k \right]
= 0,
$$
it can be shown that
\begin{align}
\uE[||\hat \bg - \bg||^2] = 
\frac{1}{K^2} \sum_{k=1}^K 
\uE \left[ Z_k\right],
	\label{EQ:C_AMSE}
\end{align}
where $Z_k = || V_{\rm max} \beta_k \bc_{m(k)} - 
||\bv_k|| \tilde \bv_k ||^2]$. 
Then, since
\begin{align}
Z_k
& = V_{\rm max}^2 \beta_k ||\bc_{m(k)}||^2 
+ ||\bv_k||^2 \cr
& \quad - 2 V_{\rm max} \beta_k ||\bv_k||
 \bc_{m(k)}^\rT \tilde \bv_k \cr
& = V_{\rm max}^2 \beta_k  L
+ ||\bv_k||^2 
 - 2 V_{\rm max} \beta_k ||\bv_k||
 \bc_{m(k)}^\rT \tilde \bv_k 
\end{align}
and $\uE[\bc_{m(k)}] = \tilde \bv_k$,
we have
\begin{align}
\uE \left[Z_k
\right] 
& = V_{\rm max}^2 \uE[\beta_k]  L
+ ||\bv_k||^2 
 - 2 V_{\rm max} \uE[\beta_k] ||\bv_k|| \cr
& =  L V_{\rm max} ||\bv_k|| - ||\bv_k||^2.
	\label{EQ:EZk}
\end{align}
Substituting \eqref{EQ:EZk}
into \eqref{EQ:C_AMSE},
we have \eqref{EQ:AMSE}.


\bibliographystyle{ieeetr}
\bibliography{ml}

\begin{thebibliography}{10}

\bibitem{Bishop06}
C.~M. Bishop, {\em Pattern Recognition and Machine Learning (Information
  Science and Statistics)}.
\newblock Berlin, Heidelberg: Springer-Verlag, 2006.

\bibitem{Goodfellow16}
I.~Goodfellow, Y.~Bengio, and A.~Courville, {\em Deep Learning}.
\newblock MIT Press, 2016.
\newblock \url{http://www.deeplearningbook.org}.

\bibitem{Verbraeken20}
J.~Verbraeken, M.~Wolting, J.~Katzy, J.~Kloppenburg, T.~Verbelen, and J.~S.
  Rellermeyer, ``A survey on distributed machine learning,'' {\em ACM Comput.
  Surv.}, vol.~53, Mar. 2020.

\bibitem{FO16}
J.~Konecn{\'y}, H.~B. McMahan, D.~Ramage, and P.~Richt{\'a}rik, ``Federated
  optimization: Distributed machine learning for on-device intelligence,'' {\em
  ArXiv}, vol.~abs/1610.02527, 2016.

\bibitem{Yang19}
Q.~Yang, Y.~Liu, T.~Chen, and Y.~Tong, ``Federated machine learning: Concept
  and applications,'' {\em ACM Trans. Intell. Syst. Technol.}, vol.~10,
  pp.~12:1--12:19, Jan. 2019.

\bibitem{Bottou18}
L.~Bottou, F.~E. Curtis, and J.~Nocedal, ``Optimization methods for large-scale
  machine learning,'' {\em SIAM Review}, vol.~60, no.~2, pp.~223--311, 2018.

\bibitem{Pokhrel20}
S.~R. {Pokhrel} and J.~{Choi}, ``Improving {TCP} performance over {WiFi} for
  {I}nternet of {V}ehicles: A federated learning approach,'' {\em IEEE Trans.
  Vehicular Technology}, vol.~69, no.~6, pp.~6798--6802, 2020.

\bibitem{Samarakoon20}
S.~{Samarakoon}, M.~{Bennis}, W.~{Saad}, and M.~{Debbah}, ``Distributed
  federated learning for ultra-reliable low-latency vehicular communications,''
  {\em IEEE Trans. Communications}, vol.~68, no.~2, pp.~1146--1159, 2020.

\bibitem{Savazzi20}
S.~{Savazzi}, M.~{Nicoli}, and V.~{Rampa}, ``Federated learning with
  cooperating devices: A consensus approach for massive {IoT} networks,'' {\em
  IEEE Internet of Things Journal}, vol.~7, no.~5, pp.~4641--4654, 2020.

\bibitem{Amiri20}
M.~M. {Amiri} and D.~{Gündüz}, ``Federated learning over wireless fading
  channels,'' {\em IEEE Trans. Wireless Communications}, vol.~19, no.~5,
  pp.~3546--3557, 2020.

\bibitem{Zhu20}
G.~{Zhu}, Y.~{Wang}, and K.~{Huang}, ``Broadband analog aggregation for
  low-latency federated edge learning,'' {\em IEEE Trans. Wireless
  Communications}, vol.~19, no.~1, pp.~491--506, 2020.

\bibitem{Yang20}
K.~{Yang}, T.~{Jiang}, Y.~{Shi}, and Z.~{Ding}, ``Federated learning via
  over-the-air computation,'' {\em IEEE Trans. Wireless Communications},
  vol.~19, no.~3, pp.~2022--2035, 2020.

\bibitem{Nazer07}
B.~{Nazer} and M.~{Gastpar}, ``Computation over multiple-access channels,''
  {\em IEEE Trans. Information Theory}, vol.~53, no.~10, pp.~3498--3516, 2007.

\bibitem{Goldenbaum13}
M.~{Goldenbaum}, H.~{Boche}, and S.~{Stańczak}, ``Harnessing interference for
  analog function computation in wireless sensor networks,'' {\em IEEE Trans.
  Signal Processing}, vol.~61, pp.~4893--4906, Oct 2013.

\bibitem{Bockelmann16}
C.~Bockelmann, N.~Pratas, H.~Nikopour, K.~Au, T.~Svensson, C.~Stefanovic,
  P.~Popovski, and A.~Dekorsy, ``Massive machine-type communications in {5G}:
  physical and {MAC}-layer solutions,'' {\em IEEE Communications Magazine},
  vol.~54, pp.~59--65, Sep 2016.

\bibitem{Ding_20Access}
J.~{Ding}, M.~{Nemati}, C.~{Ranaweera}, and J.~{Choi}, ``{IoT} connectivity
  technologies and applications: A survey,'' {\em IEEE Access}, vol.~8,
  pp.~67646--67673, 2020.

\bibitem{Choi20WCL}
J.~{Choi} and S.~R. {Pokhrel}, ``Federated learning with multichannel
  {ALOHA},'' {\em IEEE Wireless Communications Letters}, vol.~9, no.~4,
  pp.~499--502, 2020.

\bibitem{Kim20}
J.~{Kim}, G.~{Lee}, S.~{Kim}, T.~{Taleb}, S.~{Choi}, and S.~{Bahk}, ``Two-step
  random access for {5G} system: Latest trends and challenges,'' {\em IEEE
  Network}, pp.~1--7, 2020.

\bibitem{Choi21}
J.~{Choi}, ``On fast retrial for two-step random access in {MTC},'' {\em IEEE
  Internet of Things J.}, vol.~8, no.~3, pp.~1428--1436, 2021.

\bibitem{Gandikota19}
V.~Gandikota, R.~K. Maity, and A.~Mazumdar, ``{vqSGD}: Vector quantized
  stochastic gradient descent,'' {\em CoRR}, vol.~abs/1911.07971, 2019.

\bibitem{Bernstein18}
J.~Bernstein, Y.-X. Wang, K.~Azizzadenesheli, and A.~Anandkumar, ``sign{SGD}:
  Compressed optimisation for non-convex problems,'' in {\em Proceedings of the
  35th International Conference on Machine Learning} (J.~Dy and A.~Krause,
  eds.), vol.~80 of {\em Proceedings of Machine Learning Research},
  (Stockholmsmässan, Stockholm Sweden), pp.~560--569, PMLR, 10--15 Jul 2018.

\bibitem{Alistarh17}
D.~Alistarh, D.~Grubic, J.~Li, R.~Tomioka, and M.~Vojnovic, ``{QSGD}:
  Communication-efficient {SGD} via gradient quantization and encoding,'' in
  {\em Advances in Neural Information Processing Systems} (I.~Guyon, U.~V.
  Luxburg, S.~Bengio, H.~Wallach, R.~Fergus, S.~Vishwanathan, and R.~Garnett,
  eds.), vol.~30, pp.~1709--1720, Curran Associates, Inc., 2017.

\bibitem{Marzetta10}
T.~L. Marzetta, ``Noncooperative cellular wireless with unlimited numbers of
  base station antennas,'' {\em IEEE Trans. Wireless Communications}, vol.~9,
  pp.~3590--3600, Nov. 2010.

\bibitem{Krizhevsky09}
A.~Krizhevsky and G.~Hinton, ``Learning multiple layers of features from tiny
  images,'' {\em Master's thesis, Department of Computer Science, University of
  Toronto}, 2009.

\end{thebibliography}

\end{document}